%% file: template.tex
\documentclass{vgtc}                          

\ifpdf
  \pdfoutput=1\relax                   
  \usepackage{graphicx}                
  \DeclareGraphicsExtensions{.pdf,.png,.jpg,.jpeg} 
\else
  \usepackage{graphicx}                
  \DeclareGraphicsExtensions{.eps}     
\fi%

\graphicspath{{figures/}{pictures/}{images/}{./}} 
\usepackage{microtype}                 
\PassOptionsToPackage{warn}{textcomp}  
\usepackage{textcomp}                  
\usepackage{mathptmx}                  
\usepackage{times}                     
\usepackage{cite}                      
\usepackage{tabu}                      
\usepackage{booktabs}                  

\usepackage{changes}
\definecolor{cb_orange}{rgb}{1.0,0.51,0.0}
\definecolor{cb_blue}{rgb}{0.22,0.49,0.72}
\definecolor{cb_green}{rgb}{0.3,0.67,0.29}
\definecolor{cb_red}{rgb}{0.89,0.1,0.11}
\definecolor{cb_purple}{rgb}{0.6, 0.31, 0.64}
\definecolor{cb_brown}{rgb}{0.6, 0.4, 0.2}
\definecolor{cb_crimson}{rgb}{0.86, 0.08, 0.24}

\newcommand{\ecancel}{\includegraphics[scale=0.16]{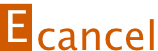}}
\newcommand{\enote}{\includegraphics[scale=0.16]{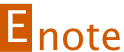}}
\newcommand{\ereport}{\includegraphics[scale=0.16]{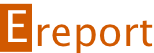}}
\newcommand{\pmountain}{\includegraphics[scale=0.16]{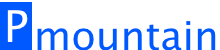}}
\newcommand{\pfireworks}{\includegraphics[scale=0.16]{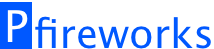}}
\newcommand{\pwedding}{\includegraphics[scale=0.16]{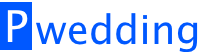}}

\usepackage{setspace}
\usepackage{lipsum}
\usepackage{etoolbox}
\AtBeginEnvironment{quote}{\singlespace\vspace{-\topsep}\small}
\AtEndEnvironment{quote}{\vspace{-\topsep}\endsinglespace}

\onlineid{0}

\vgtccategory{Research}

\vgtcinsertpkg

\title{XR Input Error Mediation for Hand-Based Input: Task and Context Influences a User's Preference}

\author{
Tica Lin\thanks{e-mail: mlin@g.harvard.edu. Work done during an internship at Meta.}\\ %
        \parbox{1.4in}{\scriptsize \centering Reality Labs Research, Meta \\ Harvard University}
\and Ben Lafreniere\thanks{e-mail: \{benlafreniere, yanx, dwigdor, mglueck\}@meta.com}\\ %
     \scriptsize Reality Labs Research, Meta %
\and Yan Xu$^\dag$\\ %
     \scriptsize Reality Labs Research, Meta %
\and Tovi Grossman\thanks{e-mail: tovi@dgp.toronto.edu}\\ %
     \parbox{1.4in}{\scriptsize \centering University of Toronto}
\and Daniel Wigdor$^\dag$\\ %
     \parbox{1.4in}{\scriptsize \centering Reality Labs Research, Meta \\ University of Toronto}
   \and Michael Glueck$^\dag$\\ %
     \scriptsize Reality Labs Research, Meta %
}

\teaser{
  \centering
  \includegraphics[width=\textwidth]{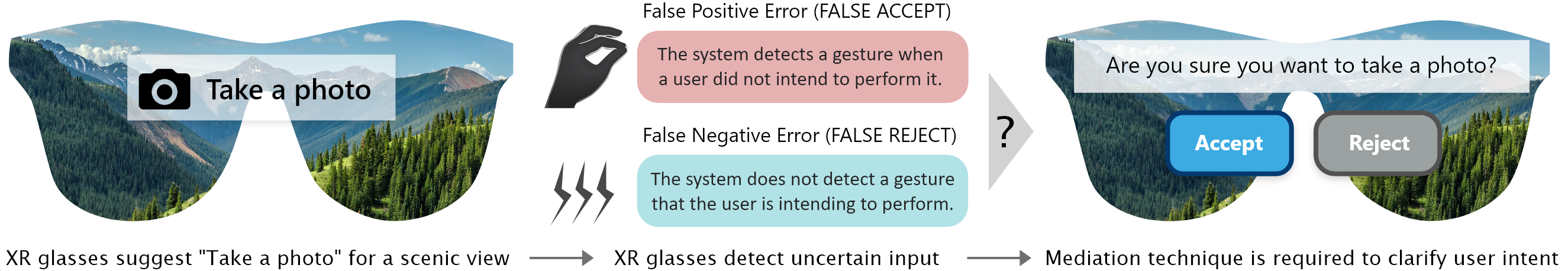}
  \caption{Error mediation techniques are designed to clarify user intentions when input errors occur, however, it is currently unclear how the user's task and context might influence their preference for error mediation designs.
  }
  \label{fig:teaser}
  \label{fig:teaser}
}

\input{docs/0_abstract}

\CCScatlist{
  \CCScatTwelve{Human-centered computing}{Human computer interaction}{Interaction paradigms}{Mixed / augmented reality}; {Interaction techniques—Gestural input}; {Empirical studies in HCI}
}

\begin{document}

\def\ND{\textsc{No Default}}
\def\DA{\textsc{Default Accept}}
\def\DR{\textsc{Default Reject}} 

\firstsection{Introduction}

\maketitle

\input{docs/1_introduction}
\input{docs/2_related_work}
\input{docs/3_web_study}
\input{docs/5_vr_study}

\input{docs/6_discussion}

\input{docs/7_conclusion}


\bibliographystyle{abbrv-doi}

\bibliography{error-mediation, other-ref}
\end{document}

%% file: docs/0_abstract.tex
\abstract{
Many XR devices use bare-hand gestures to reduce the need for handheld controllers. Such gestures, however, lead to false positive and false negative recognition errors, which detract from the user experience. While mediation techniques enable users to overcome recognition errors by clarifying their intentions via UI elements, little research has explored how mediation techniques should be designed in XR and how a user’s task and context may impact their design preferences. 
This research presents empirical studies about the impact of user perceived error costs on users' preferences for three mediation technique designs, under different simulated scenarios that were inspired by real-life tasks. Based on a large-scale crowd-sourced survey and an immersive VR-based user study, our results suggest that the varying contexts within each task type can impact users’ perceived error costs, leading to different preferred mediation techniques. We further discuss the study implications of these results on future XR interaction design. 
}

%% file: docs/1_introduction.tex


As users increasingly rely on the information and social connections available from digital technologies throughout their daily life, many envision that extended reality (XR) will eliminate the boundary between the physical and digital worlds, enabling users to interact with bare-hand gestures. However, an open challenge remains that input recognition uncertainty exists within probabilistic systems, which can lead to detection errors ~\cite{schwarz_framework_2010}. Specifically, Peacock et al. argue that in XR environments, input recognition errors are still prevalent and "can significantly degrade the usability and experience of a system" ~\cite{PeacockGaze}. Two types of errors must be considered: false accept errors occur when a system detects a gesture that a user did not intend to perform; false reject errors occur when a system does not detect a gesture that a user did intend to perform. It is therefore important to understand how input detection errors are perceived under different scenarios and to develop solutions to mitigate their negative impact on user experience.

Error mediation techniques are one way that interface design can support users in discovering and correcting system ambiguity ~\cite{mankoff2000providing}. 
Several techniques have been designed to address unclear input within context-aware systems ~\cite{bolt1980put, dey2005designing,irawati2006move,maues2013keep, olwal2003senseshapes, piumsomboon2014grasp,williamson2006continuous}.
However, previously proposed error mediation techniques for context-aware systems have not considered the impact of task and context on a user's preferences for mediation techniques in ubiquitous contexts.
A user's current task may impact the implications of a recognition error, which we term the "error cost".  
To better understand how to design error mediation for hand-based input in XR environments, which adapts to diverse XR use cases, this paper aims to answer two research questions: \textit{``How do different real-life tasks and scenarios impact users' perceived costs of input recognition errors?''} and \textit{``How do perceived error costs impact the users' choice of mediation techniques?''}.

In service of these research questions, we first characterized the consequences of input recognition errors using a cost matrix. In a large-scale web study (Sec.~\ref{sec:webstudy}), we designed and tested six simulated scenarios based on real-world tasks with varying error consequences, including photo-taking and email-sending. 
We asked users to correct input errors using three distinct mediation techniques for each scenario and rank their preferences for the techniques. The techniques accepted or rejected input by default using a timed prompt for user correction (\DA{} and \DR{}), or prompted the user for confirmation without a default action (\ND{}).

Our study results (Sec.~\ref{sec:webstudyresults}) suggested that different tasks evoked different perceived costs of false accept (FA) and false reject (FR) errors, and that scenarios with varying temporal and social stress impacted one's choice of preferred mediation technique. 
The perceived cost of FA errors was also found to be the driving force behind one's preferred mediation technique. 
For scenarios with low FA costs, \DA{} was preferred for its ease of use, while for scenarios with high FA cost, users preferred \ND{} and \DR{} due to their abilities to prevent potential FA errors.

Building upon the results of the web-based study, we designed a bare-hand gestural input mediation interface and used it within a VR-based user study (Sec.~\ref{sec:vrstudy}). We 
recreated the selected tasks and scenarios from the web study using 360 videos to validate the impact of perceived error costs on the preferred mediation techniques in an immersive environment. 
We found that preferences for mediation techniques were consistent across the web survey and VR study.

This work contributes to the design of appropriate XR error mediation techniques from users' perspectives in two ways. First, we present a crowd-sourced web study that sought to understand the impact of scenarios on user perceived error costs and preferred mediation techniques in a controlled, 2D web environment. 
The study provided concrete evidence that task context impacted one's choice of mediation techniques.
Second, we followed up with an immersive VR user study to replicate the web survey scenarios using a gestural-based mediation interface prototype and validated the impact of task and context on user preferences.

%% file: docs/2_related_work.tex

\section{Related Work}

The present research was motivated by prior literature on error handling strategies, error mediation techniques for gestural input, and techniques to address uncertain input in context-aware systems.

\subsection{Error Handling in Recognition-Based Interfaces}

The design of techniques to overcome input errors has been an essential cornerstone of recognition-based input systems. As identified by Mankoff and Abowd~\cite{mangkoejomr_error_nodate}, there are three ways that errors can be handled in recognition-based interfaces: employing error reduction strategies, improving error discovery, or using error correction techniques. Research by Bourguet~\cite{bourguet_towards_2006} later refined this categorization to focus on the role that both the user and the system have in error handling. Our research further extends these error handling strategies by focusing on the important role that users' error cost perceptions have on error mediation strategies.

\subsection{Overcoming Gesture-based Errors in User Interfaces}
In gesture-based interaction, false positives are often caused by the "Midas Touch" problem, where natural movements are falsely recognized as input gestures by a system~\cite{jacob1990you}. This problem has been the inspiration for a variety of techniques, such as using dwelling~\cite{jacob1990you}, clutching ~\cite{istance2008snap,vogel2005distant}, confirmation gestures~\cite{mohan2018dualgaze,hinckley2005design}, supplementary input modalities~\cite{wang2020blyncsync}, or recognizing user intent~\cite{PeacockGaze, SendhilnathanDetecting, schwarz2014combining, zhangrids}. 

To mitigate the Midas Touch problem, the use of visual feedback has been proposed ~\cite{dey2005designing, schwarz_framework_2010}.
Ripples, for example, categorized the ambiguous input that can occur when interacting with a multi-touch tabletop and presented water-like "ripples" to visualize the current input state or transition~\cite{wigdor_ripples_2009}. OctoPocus helped users learn marking-menu style gestures~\cite{bau_octopocus_2008}. It visualized the n-best gestures that a user may be performing using visual feedforward as the user continued to perform their gesture. 
These techniques enable users to understand when input errors occur (Ripples) and provide feedback to prevent users from making erroneous input (OctoPocus). 


\subsection{Error Mediation in Context-Aware Systems}
Context-aware systems use input from a user’s environment and activities to make inferences about their intentions and generate appropriate responses. From intelligent assistants to ubiquitous computing and mixed-reality devices, the necessity to incorporate human control in the decision-making processes of an automation system has been frequently studied in research on mixed-initiative systems~\cite{horvitz_principles_1999} and human-AI interaction~\cite{amershi_guidelines_2019}. To support this design perspective, previous work has highlighted how users with high degrees of agency can have higher degrees of user satisfaction, even when automation system accuracy is low~\cite{roy_automation_2019}.  

Similar to recognition-based input systems, inherent uncertainties exist in context-aware systems. However, with more complex task types and real-world scenarios, user acceptability towards system errors has been found to differ by context~\cite{kay_how_2015, ben2021falsepositives}. As such, a single mediation technique may not be applicable for a range of contexts. Providing adaptive interfaces that can engage users when input clarification is needed is thus necessary~\cite{horvitz_uncertainty_1999, jonker_chi20_2020}. 

Previous projects have proposed the use of different mediation techniques for different application contexts. For example, mixed-initiative systems like LookOut adjusted the default system actions of a messaging and scheduling system based on the level of confidence LookOut had about a message's content and context~\cite{horvitz_uncertainty_1999}. Dey et al. developed two interactive error-handling solutions for an in-out board and an office reminder system, with different default choices being implemented to support error handling~\cite{dey2005designing}. These default choices considered the expected utility of the tasks to be performed, so one accepted the results from an identity recognizer by default, whereas the other took no action and marked reminders as “still pending” by default. Other systems used n-best lists~\cite{maues2013keep} or explicitly prompted users for input clarification~\cite{mankoff_interaction_2006} when uncertainty occurred. Depending on a user’s tasks and context, a different default action may be selected and the user may be alerted that they need to make a correction, including “Default Accept” (e.g., carry out the action if the user does not correct it), “Default Reject” (e.g., dismiss an action if the user does not correct it), or “No Default” (e.g., provide an n-best list of options to select from). However, it is unclear which default strategy should be implemented under different contexts. Our research addresses this gap by investigating the impact of task context on users' perceptions of error costs and preferred mediation strategies.

%% file: docs/3_web_study.tex

\section{Large-Scale Web Survey}
\label{sec:webstudy}
Inspired by existing work~\cite{horvitz_uncertainty_1999,dey2005designing,maues2013keep}, where different mediation approaches were designed for different context-aware systems, our study investigated the impact of task context and error costs on users' preferred mediation strategies under varying XR contexts. 
We utilize two XR tasks, each with three real-world scenarios, to trigger different perceived error costs from participants.
A web-based survey with interactive prototypes was designed to simulate six selected real-world scenarios and collect responses from a large variety of participants,
similar to the approach used by Kay et al. ~\cite{kay_how_2015}. While a study with actual implementations that immerse participants in the scenarios may elicit more realistic feedback, such an evaluation would be difficult to perform at scale. The web-based methodology allowed us to collect initial perspectives from a large number of respondents (N=224) and is consistent with the Scenario-Based Design approach ~\cite{carroll2003making} which “evoke effective reflection in a way that addresses some of the most difficult properties of design” ~\cite{carroll2000}.
In Sec.~\ref{sec:vrstudy}, we verify and replicate the results in a VR-based study.

\subsection{Derivation of Study Tasks}
\label{sec:design_exploration}
Due to the complex and exploratory nature of real-world XR scenarios, it is difficult to accurately define and understand all factors related to error perception in a single study. Instead, our goal was to derive a range of task and context combinations and probe user perceptions under varying scenarios. To do so, we conducted a series of internal interviews and brainstorming sessions with three experienced XR designers and researchers from our institution to help formulate representative tasks and contexts for our study that would have varying error costs. 

An important observation  made during these sessions was that temporal and social factors played an important role in dictating errors costs. For instance, FR errors would be particularly important in a task with high temporal stress (e.g. calling to report a crime) whereas FA errors could be particularly important in a task with high social stress (e.g. accidentally sending an email to the wrong person). By considering these factors, we could classify tasks within a \textbf{two-by-two conceptual matrix of error costs} that consisted of the type of error (i.e., false accept or false reject) and the intensity of the error costs (i.e., low or high). We then selected two contrasting tasks on the error cost matrix, including a \textbf{photo-taking} task (i.e., high FR cost) and an \textbf{email-sending} task (i.e., high FA cost). Each task was then designed with three scenarios that had distinct levels of social and temporal stress, resulting in six XR scenarios (Table~\ref{table:1}).

 \begin{table}[t!]
    \centering
       \scriptsize
        \begin{tabular}{ p{7em} c| c c } 
         Task & Scenario & \textit{Social Stress} & \textit{Temporal Stress}  \\ 
         \hline
         \textbf{Photo Taking} & {\pmountain{}} & low & low \\ 
         & \pfireworks{} & low & high \\
         & \pwedding{} & high & high \\
         \hline
         \textbf{Email Sending} & {\enote{}} & low & low \\ 
         & \ereport{} & low & high \\
         & \ecancel{} & high & high \\
         \hline
        
        \end{tabular}
    \caption{The six XR scenarios used in the web survey varied the level of social and temporal stress}
    \label{table:1}
    \vspace{-6mm}
\end{table}



\subsubsection{Photo Taking}
Generally speaking, photo-taking has a higher cost of FR errors than FA errors, as not taking a desired photo can be more costly than accidentally taking a photo. 
However, the consideration of errors becomes more nuanced when photos are taken by always-on XR glasses, where errors can lead to varying privacy and social consequences. 
Our XR photo-taking tasks assumed an XR glasses experience where an application could identify a user’s context and suggest that a photo could be taken, and while taking a photo, a camera sound was played to prevent privacy violations. 
To capture the different FA and FR error perceptions associated with photo-taking, 
we designed three scenarios with varying levels of social and temporal stress (Table~\ref{table:1}), including taking scenic pictures of the mountains (\pmountain{}; low social and low temporal stress),  taking pictures of fireworks (\pfireworks{}; low social and high temporal stress), and taking wedding party photos (\pwedding{}; high social and high temporal stress).
Each scenario was introduced with a short narrative and followed by a description of the consequences when FA and FR errors occurred. The complete task descriptions are provided in Appendix A. The respondents then reported their frustration level when FA and FR errors occurred on 7-point Likert scales.


\subsubsection{Email Sending}
When sending an email, it is typically more costly to unintentionally send an email than not send an email due to the irreversible, bilateral nature of the communication. 
Considering the prevalent use of email as a communication tool in the workplace, we asked respondents to imagine that they were using a productivity application on their XR glasses that streamlined their calendars, meetings, and file management. The application would provide suggestions to send emails based on one’s calendar activities. We designed three scenarios with different levels of stress, including emailing a meeting note to one’s team (\enote{}; low social and low temporal stress), emailing a report by an urgent, upcoming deadline (\ereport{};  high temporal stress), and sending an emergency email to cancel a meeting while driving (\ecancel{}; high temporal and high social stress).
The complete task descriptions are provided in Appendix A.


\begin{figure}[t!]
    \centering
    \includegraphics[width=0.48\textwidth]{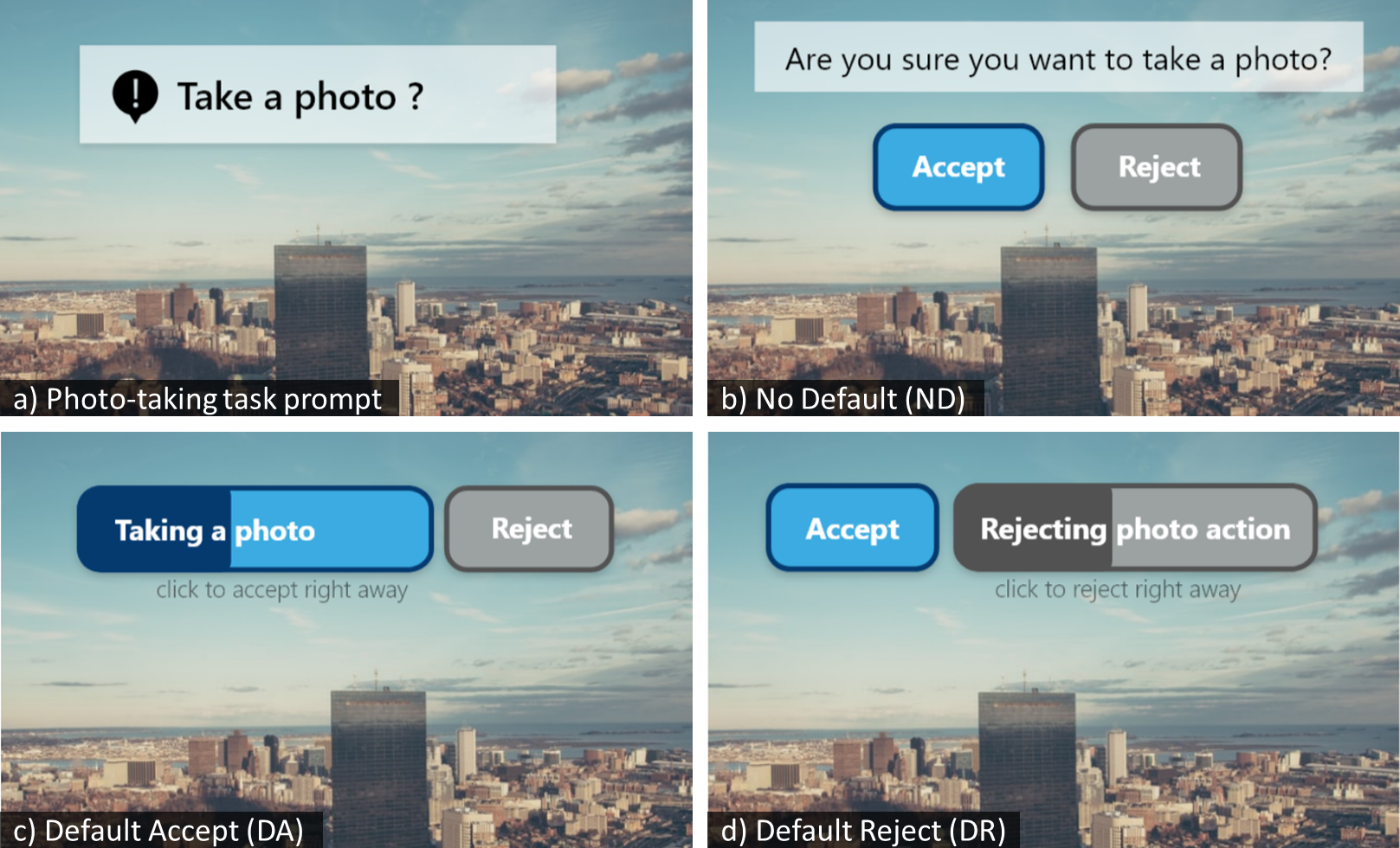}
    \vspace{-6mm}
    \caption{The photo-taking task prompt and the three interactive mediation interfaces used in the web survey.} 
    \label{fig:task_ui}
    \vspace{-4mm}
\end{figure}

\subsection{Mediation Techniques}
Mediation techniques were designed to clarify user intent with interfaces when a system detected uncertain user input.
To understand users' preferred mediation techniques under different scenarios, three representative mediation strategies were implemented.
As shown in Fig.~\ref{fig:task_ui}, survey respondents were first prompted with the task prompt, i.e., “Take a Photo?” in the photo-taking or “Send the Email?” in the email-sending scenarios. 
Then, they interacted with each mediation technique using a 2D interactive prototype and compared each technique under different XR scenarios.

\textbf{No Default} (ND) was set to wait until the respondent confirmed their intention. The interface displayed an explicit confirmation dialog, e.g., “Are you sure you want to take a photo?”, an accept button below and to the left of the dialog, and a reject button below and to the right of the dialog (Fig.~\ref{fig:task_ui}-2). Respondents needed to click on the accept or reject button to reply to the suggestion. This technique was used because it was akin to modal dialog boxes that are found in interfaces today, which block further action until a user has provided their input.

\textbf{Default Accept} (DA) was set to automatically accept input after a period of time if there was no further clarification from the user. The interface displayed an elongated accept button on the left and a normal reject button on the right (Fig.~\ref{fig:task_ui}-3). The extended accept button consisted of a progress bar and descriptive text indicating the ongoing action by the XR glasses, such as “Taking a Photo”. At the start of the task, the progress bar began to advance to the right. To accept the suggestion, the respondent could let the bar advance all the way to the right or they could click on the accept button. To reject the suggestion, the respondent would need to click on the reject button before the progress bar completely advanced to the right. This technique was similar to applications that accept the first input that is detected but allow a period of time for correction, such as the in-out board developed by Dey and Mankoff~\cite{dey2005designing} or the undo button that is displayed shortly after one sends an email for immediate correction~\cite{gmail_undo_2021}. 

\textbf{Default Reject} (DR) was similar to the \DA{} technique, however, the progress bar was part of the reject button rather than the accept button (Fig.~\ref{fig:task_ui}-4). To reject the suggestion, the respondent could let the bar automatically advance all the way to the right or they could click on the reject button. To accept the suggestion, the user needed to click on the accept button before the progress bar completely advanced to the right.  This technique was used because it was similar to the use of dwelling over an interface element when using a mouse, performing a gesture, or gazing for XR input~\cite{mixedrealitytoolkit_2021}, i.e., a system will reject one’s input if insufficient clarification is provided.

\begin{figure*}[t!]
    \centering
    \includegraphics[width=0.9\textwidth]{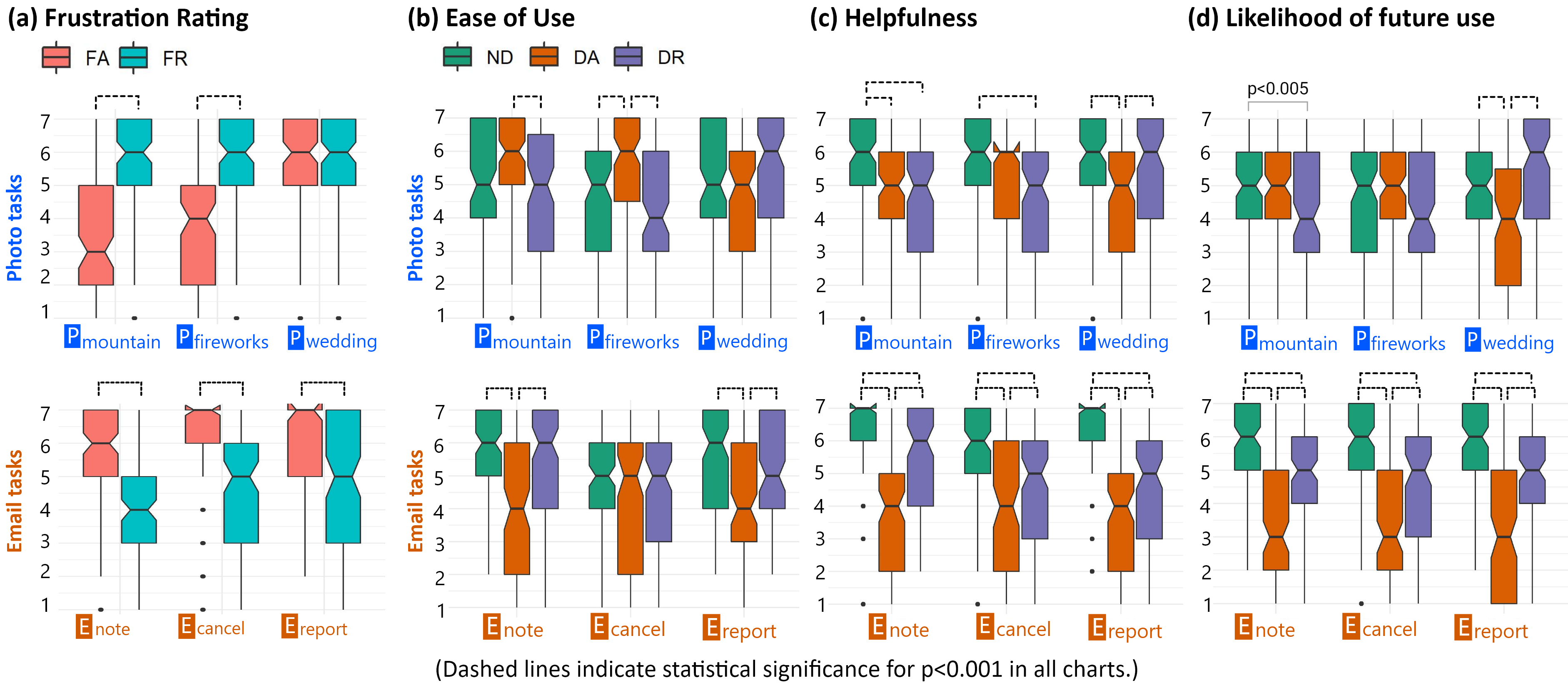}
    \vspace{-3mm}
    \caption{Box plots of the median responses for (a) Frustration and (b-d) TAM ratings for each task. 
    (a) All three photo-taking tasks received high frustration ratings for FR. The three email-sending tasks received high frustration for FA. (b) DA was rated easiest to use for scenarios with lower FA costs (\pmountain{}, \pfireworks{}), and received the lowest ratings for all scenarios with high FA costs (\pwedding{}, email-sending). (c) ND was perceived to be most helpful to clarify the input for both tasks. (d) Preferred mediation techniques vary based on scenarios and tasks.
    }
    \label{fig:rating}
     \vspace{-3mm}
\end{figure*}

\subsection{Respondents and Procedure}
The study respondents were directed to
a Qualtrics~\cite{qualtrics_2021} survey and were required to use a desktop or laptop with a mouse to complete the survey. No XR hardware was required. 
The complete surveys are provided in Appendix A.

\textbf{Respondents.} 224 respondents completed the survey in English. 112 respondents completed the photo-taking survey and 112 respondents completed the email-sending survey. All respondents were recruited using Amazon Mechanical Turk in the US (with HIT approval rate $\geq$ 98\% and approval score $\geq$ 1000). 
To ensure response reliability, three attention check questions were embedded in the survey. After removing responses that failed the attention check questions, in the photo-taking survey, 99 out of 112 responses (88\% success rate) were included in the study analysis (M=63, F=35, Non-binary=1; Age: 18 to above 65 with 78\% between 18 to 45). 
In the email-sending survey, 105 out of 112 responses (94\%) were included in the study analysis
(M=67, F=37, Prefer not to say=1; Age: 18 to above 65 with 83\% between 18 to 45).
Most respondents had
occasional or no experience using AR/VR headsets in the photo-taking (89\%) and email-sending survey (90\%) populations.

\textbf{Procedure.} The survey had two sections. The first section introduced three mediation techniques to respondents using interactive prototypes (Fig.~\ref{fig:task_ui}). The prototypes were implemented using Adobe XD~\cite{adobexd_2021} and embedded in the survey. Respondents were asked to interact with the prototypes to accept and reject the XR glasses’ recommendation about uncertain input. This allowed respondents to experience both false reject and false accept errors. In the second section of the survey, three scenarios were presented using images and a narrative to engage respondents in the scenario. Respondents first rated the perceived cost of consequences when different types of errors occurred and then evaluated each technique's acceptability for each scenario.
After they worked through all three scenarios, Likert-scale and open-ended response questions were used to elicit feedback about the three mediation techniques. 
The survey took up to 40 minutes to complete and respondents were compensated \$10 USD for completing the survey.

\subsection{Measures and Analysis}
To understand the perceived error costs, respondents rated their level of frustration if the suggestion was falsely accepted (FA) or rejected (FR) for each scenario on a 7-point Likert scale. Example questions included \textit{``I would feel frustrated if my AR glasses wrongly took a photo when I didn't want them to''} (FA) and \textit{``I would feel frustrated if my AR glasses did not take a photo when I wanted to''} (FR).
To understand the acceptability of each mediation technique, we adopted questions from Technology Acceptance Model~\cite{davis1989perceived}.
For each technique under each scenario, respondents rated on a 7-point Likert scale on ease of use (\textit{``I feel that this technique would be easy to use without interfering with my current task''}), helpfulness (\textit{``I feel that this technique would be helpful to clarify
my intentions''}), and intent to use (\textit{``I would be more likely to use my AR glasses with
this technique when detection errors happen''}).
Respondents also provided a ranking of all three techniques for each scenario.

For non-parametric rating data, we first converted the data into normalized data distributions using the Aligned Rank Transform Procedure~\cite{higgins1990aligned,wobbrock2011aligned}.
Factorial ANOVAs were then performed using the aligned data. Post-hoc analyses for main effects were conducted using Bonferroni-corrected pairwise comparisons. When post-hoc analyses were needed to analyze interaction effects, the ART-C procedure was used to align the data to reduce Type I errors ~\cite{elkin2021aligned}.
Each factorial ANOVA followed a 3 x 2 design, i.e., Scenario: 1, 2, 3; ErrorType: FA (false accept) and FR (false reject), or 3 x 3 design, i.e., Scenario: 1, 2, 3; Technique: ND (\ND{}), DA (\DA{}), DR (\DR{}) depending on the dependant variable that was being measured.

Qualitative data was collected via open-ended questions about respondents’ preferred techniques and rationales. Random sampling was used to select 20 responses from each survey due to the large number of respondents. All responses were first assigned a number from 1 to 10 in the order of response time, and responses were drawn from three randomly selected groups (3, 5, 8), while skipping incomplete responses. A total of 40 responses were qualitatively encoded and analyzed by two coders.


\vspace{-2mm}
\section{Web Survey Results}
\label{sec:webstudyresults}
We first present the statistical and qualitative analysis results for both surveys, and synthesize our findings on the impact of XR tasks and contexts to user's error perceptions and mediation preferences.

\subsection{Error Perception - Frustration Rating Results}
In the photo-taking survey, a significant main effect of Scenario (\textit{F}(1.78, 174.29) = 54.45, \textit{p} $<$ 0.001, $\eta_p^2$ = 0.36) and ErrorType
(\textit{F}(1,98) = 135.04, \textit{p} $<$ 0.001, $\eta_p^2$ = 0.58) were found for frustration. An interaction between Scenario and ErrorType 
was also found (\textit{F}(1.89, 184.73) = 48.55, \textit{p} $<$ 0.001, $\eta_p^2$ = 0.33).
Post-hoc comparisons revealed that it was more frustrating when FR errors occurred in the \pmountain{} and \pfireworks{} scenarios, however, in the \pwedding{} scenario, there was no difference in frustration (Fig.~\ref{fig:rating}a). 
 For email-sending, a significant main effect of Scenario (\textit{F}(2, 208) = 23.63, \textit{p} $<$ 0.001, $\eta_p^2$ = 0.19) and ErrorType 
 (\textit{F}(2,208) = 129.04, \textit{p} $<$ 0.001, $\eta_p^2$ = 0.55) were also found for frustration. The ANOVA revealed an interaction between Scenario and ErrorType
 (\textit{F}(1.88, 1195.29) = 7.99, \textit{p} $<$ 0.001, $\eta_p^2$ = 0.07) and post-hoc comparisons showed that it was more frustrating when FA errors occurred than FR errors in the \enote{}, \ecancel{}, and \ereport{} scenarios (Fig.~\ref{fig:rating}a). 
 
 These results suggest that \textbf{tasks have impact on users' error perception} and \textbf{even for the same task, the perceived error costs can be different under varying task scenarios,}.
For photo-taking, FA errors were perceived as being less detrimental during scenarios with varying time pressure (\pmountain{} and \pfireworks{}) compared to those with social pressure (\pwedding{}), but FR errors were perceived to be detrimental in all scenarios. However, in the email-sending tasks, FA errors were more detrimental than FR errors, irrespective of scenario demands. 

\subsection{Technology Acceptance Model Ratings}
Figure~\ref{fig:rating}b shows the ratings for each mediation technique based on their ease of use, helpfulness, and likelihood of future use.

\textbf{(i) Ease of Use.}
With the photo-taking tasks, no significant main effects of Scenario (\textit{F}(2, 196) = 2.34, \textit{p} = 0.10, $\eta_p^2$ = 0.02) or Technique (\textit{F}(2,196) = 2.65, \textit{p} = 0.07, $\eta_p^2$ = 0.03) were found, however there was an interaction between Scenario and Technique (\textit{F}(3.41, 334.35) = 13.09, \textit{p} $<$ 0.001, $\eta_p^2$ = 0.12). Post-hoc comparisons found that for the \pmountain{} scenario, the ease of use ratings for DA were higher than the DR technique. For the \pfireworks{} scenario, ratings of DA were higher than the ND and DR techniques. This suggests that the \textbf{perceived ease of use with DA was higher than DR in scenarios with higher cost of FR (i.e., \pmountain{} and \pfireworks{}) and higher than ND under increased time pressure (i.e., \pfireworks{}).}

With the email-sending tasks, an ANOVA revealed significant main effects of Scenario (\textit{F}(1.18, 189.14) = 3.34, \textit{p} $<$ 0.05, $\eta_p^2$ = 0.03) and Technique (\textit{F}(2,208) = 13.57, \textit{p} $<$ 0.001, $\eta_p^2$ = 0.12), and an interaction between Scenario and Technique 
(\textit{F}(4, 416) = 9.70, \textit{p} $<$ 0.001, $\eta_p^2$ = 0.09). 
Post-hoc tests revealed that for the \enote{} scenario, ease of use ratings of ND were higher than DA and ratings of DR were higher than DA. For the \ereport{} scenario, the ease of use ratings of ND and DR were higher than DA. \textbf{Across all email scenarios, DA received lower ease of use ratings, as it required explicit input to reject the FA errors compared to DR and ND.} In the \ecancel{} scenario, there may have been too many consequences for one to consider, resulting in all techniques being inadequate.

\textbf{(ii) Helpfulness.}
With the photo-taking tasks, a significant main effect of Technique (\textit{F}(2, 196) = 13.72, \textit{p} $<$ 0.001, $\eta_p^2$ = 0.12) was found, however no main effect of Scenario was found (\textit{F}(1.89, 185.33) = 1.78, \textit{p} = 0.17, $\eta_p^2$ = 0.02). There was also an interaction between Scenario and Technique (\textit{F}(3.19, 312.54) = 10.65, \textit{p} $<$ 0.001, $\eta_p^2$ = 0.10). Post-hoc comparisons found that for the \pmountain{} scenario, helpfulness ratings of ND were higher than DA and DR. For the \pfireworks{} scenario, ratings of ND were higher than DR. For the \pwedding{} scenario, ratings of ND and DR were higher than DA. This suggests respondents perceived \textbf{ND to be more helpful to clarify their intent in general, and DA to be the least helpful}, maybe due to the feeling of control provided by ND and the desire to avoid FA errors. Depending on the stress introduced to the scenario, the perceived helpfulness of DA increased (\pfireworks{}) or decreased (\pwedding{}).

With the email-sending tasks, the ANOVA revealed a significant main effect of Scenario (\textit{F}(2, 208) = 5.40, \textit{p} $<$ 0.005, $\eta_p^2$ = 0.05) and Technique (\textit{F}(2,208) = 81.43, \textit{p} $<$ 0.001, $\eta_p^2$ = 0.44). Pairwise comparisons found that helpfulness ratings were higher in the \enote{} compared to the \ereport{} Scenario, perhaps because they did not induce great feelings of stress or worry if an incorrect action were to be taken. Pairwise comparisons also revealed that ratings of ND were higher than DR, and DR were higher than DA. The ANOVA did not reveal an interaction between Scenario and Technique (\textit{F}(3.46, 359.98) = 2.12, \textit{p} = 0.09, $\eta_p^2$ = 0.02). These results suggest that \textbf{ND was the most helpful}, perhaps because it mimicked the common interfaces used today. Interestingly, the DR was perceived as more helpful across all scenarios compared to the DA. This suggests that although the time and social pressures varied across all tasks, \textbf{respondents were more willing to repeat their actions (i.e., DR or ND) than try to determine how to recover from them (i.e., DA).}

\textbf{(iii) Likelihood of Future Use.}
When asked if they would be likely to use their XR glasses in the future with the photo-taking tasks, a significant main effect of Scenario was found (\textit{F}(2, 196) = 3.88, \textit{p} $<$ 0.05, $\eta_p^2$ = 0.04), however, no main effect of Technique was found (\textit{F}(2,196) = 1.55, \textit{p} = 0.21, $\eta_p^2$ = 0.02). The ANOVA did reveal an interaction between Scenario and Technique (\textit{F}(3.22, 315.43) = 14.55, \textit{p} $<$ 0.001, $\eta_p^2$ = 0.13). Post-hoc tests found that for the \pmountain{} scenario, respondents would be more willing to use ND than DR technique. For the \pwedding{} scenario, respondents were more willing to use ND than DA, and DR than DA. None of the other paired t-tests revealed significant results (\textit{p} $>$ 0.05). This suggests that \textbf{technique preference was influenced more by the scenario than the task (photo), and clearer preferences emerged as more social stress was introduced into a scenario (\pwedding{}).}

With the email-sending tasks, a significant main effect of Technique (\textit{F}(2,208) = 66.36, \textit{p} $<$ 0.001, $\eta_p^2$ = 0.39) was found. Post-hoc comparisons revealed that respondents were more willing to use ND compared to DA and DR, and DR compared to DA. Interestingly, it appeared that \textbf{respondents were more willing to use existing techniques that they would have to always provide input to (i.e., ND) rather than novel techniques where there was some uncertainty about the interaction required (i.e., DA and DR)}. No main effect was found for Scenario (\textit{F}(1.88, 195.79) = 1.04, \textit{p} = 0.35, $\eta_p^2$ = 0.01) and there was no interaction between Scenario and Technique (\textit{F}(3.69, 384.14) = 2.22, \textit{p} = 0.07, $\eta_p^2$ = 0.02).

\begin{figure}[t!]
    \centering
    \includegraphics[width=0.48\textwidth]{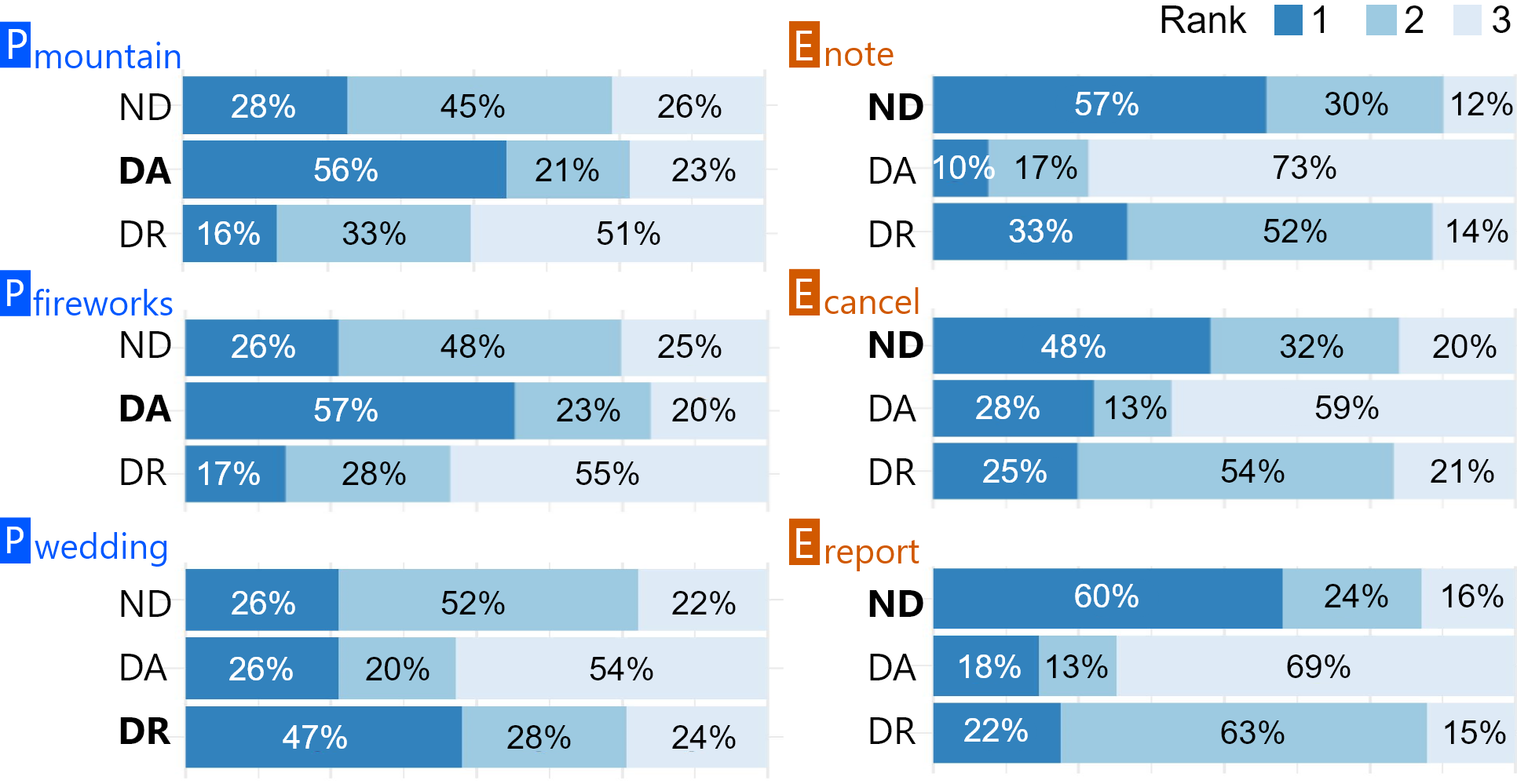}
    \vspace{-6mm}
    \caption{The mediation technique rankings for all six scenarios. The most preferred technique is shown with bold font.} 
    \label{fig:rank1}
    \vspace{-6mm}
\end{figure}

\subsection{Technique Rankings}
When asked which technique they preferred for each scenario (Fig. ~\ref{fig:rank1}), most respondents preferred the DA technique in the \pmountain{} (56\%) and \pfireworks{} (57\%) scenarios, but in the \pwedding{} scenario,  47\% 
preferred the DR technique. A Friedman’s ANOVA revealed significant differences in the most preferred technique across all scenarios ($\chi^2$(2) = 13.30, \textit{p} $<$ 0.001, \textit{W} = 0.07). Wilcoxon signed-rank post-hoc tests with a Bonferroni correction found that more respondents preferred DA in the \pfireworks{} scenario than in the \pwedding{} scenario (\textit{Z} = -3.18, \textit{p} $<$ 0.001) and preferred DA during the \pmountain{} scenario than during the \pwedding{} scenario (\textit{Z} = -3.50, \textit{p} $<$ 0.001). There was no significant difference in the number of respondents who picked DA as the most preferred in the \pmountain{} or \pfireworks{} scenario (\textit{Z} = -0.39, \textit{p} = 0.78). This suggests that when the perceived FA cost increased in the \pwedding{} scenario, the DA technique was inadequate, demonstrating that \textbf{FA was the driving force behind respondents' preferred mediation technique.}

For email-sending, respondents preferred the ND technique (i.e, 57\% in \enote{}, 48\% in \ecancel{}, and 60\% in \ereport{}), however, a Friedman’s ANOVA revealed no significant differences ($\chi^2$(2) = 5.04, \textit{p} = 0.08, \textit{W} = 0.02). These results complement those found for helpfulness and ease of use, which showed clear preferences for the ND and DR techniques, but not across specific scenarios.

\subsection{Qualitative Analysis}

To complement the quantitative results,
we analyzed the most preferred technique and rationales from 20 respondents for each survey.

\textbf{Photo-Taking Scenarios.}
In the \pmountain{} scenario, the majority of respondents (13 out of 20) preferred the DA technique due to the \textit{``low cost of FA''} (N=7) and \textit{``no input required to accept''} (N=6).
In the \pfireworks{} scenario, the majority of respondents (14) preferred DA to \textit{``avoid FR''} (10) and due to the \textit{``low cost of FA''} (9). 
In the \pwedding{} scenario, most respondents (12) preferred DR to \textit{``avoid FA''} (9), \textit{``no input required to reject''} (8), and \textit{``avoid inappropriate real-world impact''} (7). 
The results showed that \textbf{the preferred mediation techniques for the photo-taking tasks were mainly decided by the scenario consequences} (e.g., avoid FA), as expressed in \textit{``it would be mortifying to have your glasses make shutter sounds at an inopportune moment,''} \textbf{and the user interaction} (e.g., no input required to accept), \textit{``I would be annoyed if I had to tell it twice.''}

\textbf{Email-Sending Scenarios.}
In the \enote{} scenario, the majority of respondents (15) preferred ND because they \textit{“feel in control”} (12), there was \textit{“no unwanted action”} (5), and \textit{“no time pressure”} (5).
In the \ecancel{} scenario, most respondents (9) preferred ND to \textit{“feel in control”} (5) and \textit{“no time pressure”} (5). 
In the \ereport{} scenario, the majority of respondents (17) preferred ND to \textit{“feel in control”} (12) and \textit{“no unwanted action”} (8). 
These results show that \textbf{during the email-sending tasks, the preferred techniques were dominated by the subjective feelings of control and stress}, e.g., \textit{``I would always want the final say and not be rushed with a timer.''} 

Interestingly, with the \ecancel{} scenario, preferences were more diverse due to different subjective prioritization that occurred in these complex situations. In high stake situations, some respondents preferred to feel in control, i.e., \textit{``I did not want to inadvertently cancel the meeting prematurely until I was able to determine whether or not someone else could step in for me and present instead''} (ND).
Conversely, some preferred to focus on the real-world tasks in the scenarios, e.g., \textit{`` This allows me to stay attentive to driving and doesn't force me to exert any effort. I can then send the email when I have a minute.''} (DR) and 
\textit{``I prefer the method that is most hands-off, because I am driving and do not want the distraction''} (DA).

\begin{figure}[t!]
    \centering
    \includegraphics[width=0.3\textwidth]{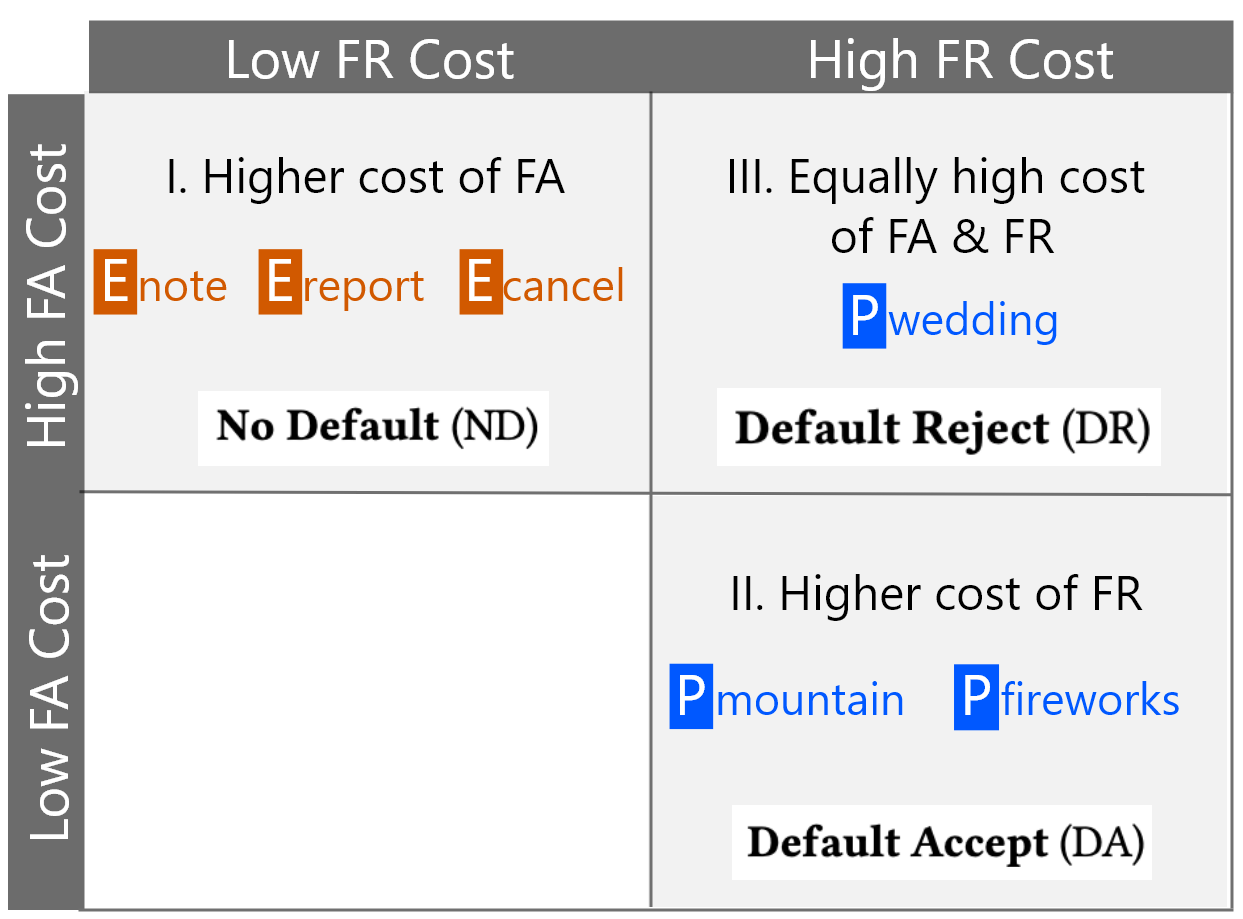}
    \vspace{-2mm}
    \caption{Three scenario types based on perceived error costs. (I) higher FA cost, (II) higher FR cost, and (III) equally high FA \& FR costs. Each type mapped to a different preferred mediation technique. } 
    \label{fig:rank2}
    \vspace{-4mm}
\end{figure}

\subsection{Discussion}
Based on quantitative and qualitative results, there was a clear preference for different mediation techniques in scenarios with varying perceived costs. We summarize three types of perceived costs identified for the different scenarios (Fig.~\ref{fig:rank2}).

(I) Type I maps to scenarios with a higher cost of FA, which includes all three scenarios in the email-sending survey (\enote{}, \ecancel{}, \ereport{}). For high FA cost scenarios, \ND{} was the most preferred. The majority of participants also ranked \DA{} as the least preferred (i.e., 73\%, 59\%, and 69\% for \enote{}, \ecancel{}, and \ereport{}, respectively). The qualitative analysis showed that in these scenarios, respondents preferred to have full control over their input clarification without time pressure to avoid FA errors. 

(II) Type II maps to scenarios with a higher cost of FR, including the \pmountain{} and \pfireworks{} photo-taking scenarios. For high FR cost scenarios,
\DA{} was the most preferred. Due to the high cost of FR, \DR{} was ranked the least preferred by the majority of respondents (51\% in \pmountain{} and 55\% in \pfireworks{}). The qualitative analysis showed that respondents preferred \DA{} in these scenarios as it worked best to avoid FR errors without user input when the cost of FA was low.

(III) Type III maps to scenarios with equally high costs of FA and FR, including the \pwedding{} scenario, for which \DR{} was the most preferred. Similar to the scenarios in Type I, \DA{} was ranked the least preferred by 54\% of respondents in \pwedding{}. However, compared to Type I, when facing a higher cost of FA and FR, the benefit of having full control in \ND{} was not as important as reducing the risk of FA errors. The qualitative analysis showed that respondents prioritized avoiding FA errors with \DR{} that did not require user input, e.g., \textit{``If the AR processor incorrectly interpreted an action of mine as a signal to take a picture, I wouldn't be required to do anything to cancel it''}. Respondents relied on manual correction for FR errors because they would already be prepared to perform a gesture, e.g., \textit{``If I was actually attempting to take a photo, I would be ready for the prompt and it wouldn't be difficult or time consuming to confirm''}.
These responses show how FA errors have a higher impact on respondents' perceptions of errors and their preferences for the mediation techniques, supporting the findings of Lafreniere et al.~\cite{ben2021falsepositives}.

Summarizing the web study results, \textbf{both task types and contexts impacted the perceived error costs}.
\textbf{User preferred mediation techniques were impacted by the perceived cost of errors and more so by FA errors than FR errors}. When the cost of FA was high, the \DA{} technique was disliked by the majority of respondents (\pwedding{}, \enote{}, \ecancel{}, \ereport{}). Therefore, \DA{} should be avoided as error mediation technique during high cost FA scenarios. Depending on the cost of FR, \ND{} could be helpful to allow full control over FA errors when the cost of FR is low. \DR{} could be used to alleviate FA errors when the cost of FR is also high. On the other hand, when the cost of FA was low (\pmountain{}, \pfireworks{}), the \DA{} technique was strongly preferred among all meditation techniques. Furthermore, as per-respondent preferences and suggestions were found in the open-ended responses, there thus seems to be a need for configurable mediation types to accommodate individual differences.  

These results are both novel and important for the future development of XR interfaces. It is the first known evidence which shows that a one-size-fits-all approach to error mediation in contextual XR interfaces may not be appropriate. Not only should error mediation techniques adapt to the current task, but they should also take into account the specific task context. In the following section, we further validate these results in an immersive XR environment and investigate appropriate input mechanisms for interacting with the error mediation techniques.

%% file: docs/5_vr_study.tex

\section{Immersive VR Study}
\label{sec:vrstudy}
Following the web study, we conducted an immersive study in VR to corroborate whether the observed preferences for different mediation defaults also occur in immersive XR scenarios. 
We replicated a subset of the web-based study, including two scenarios from the photo-taking, and implemented bare-hand gestural interfaces for the three mediation techniques. Implementation details are provided in Appendix B.

\begin{figure}[t!]
    \centering
    \includegraphics[width=0.48\textwidth]{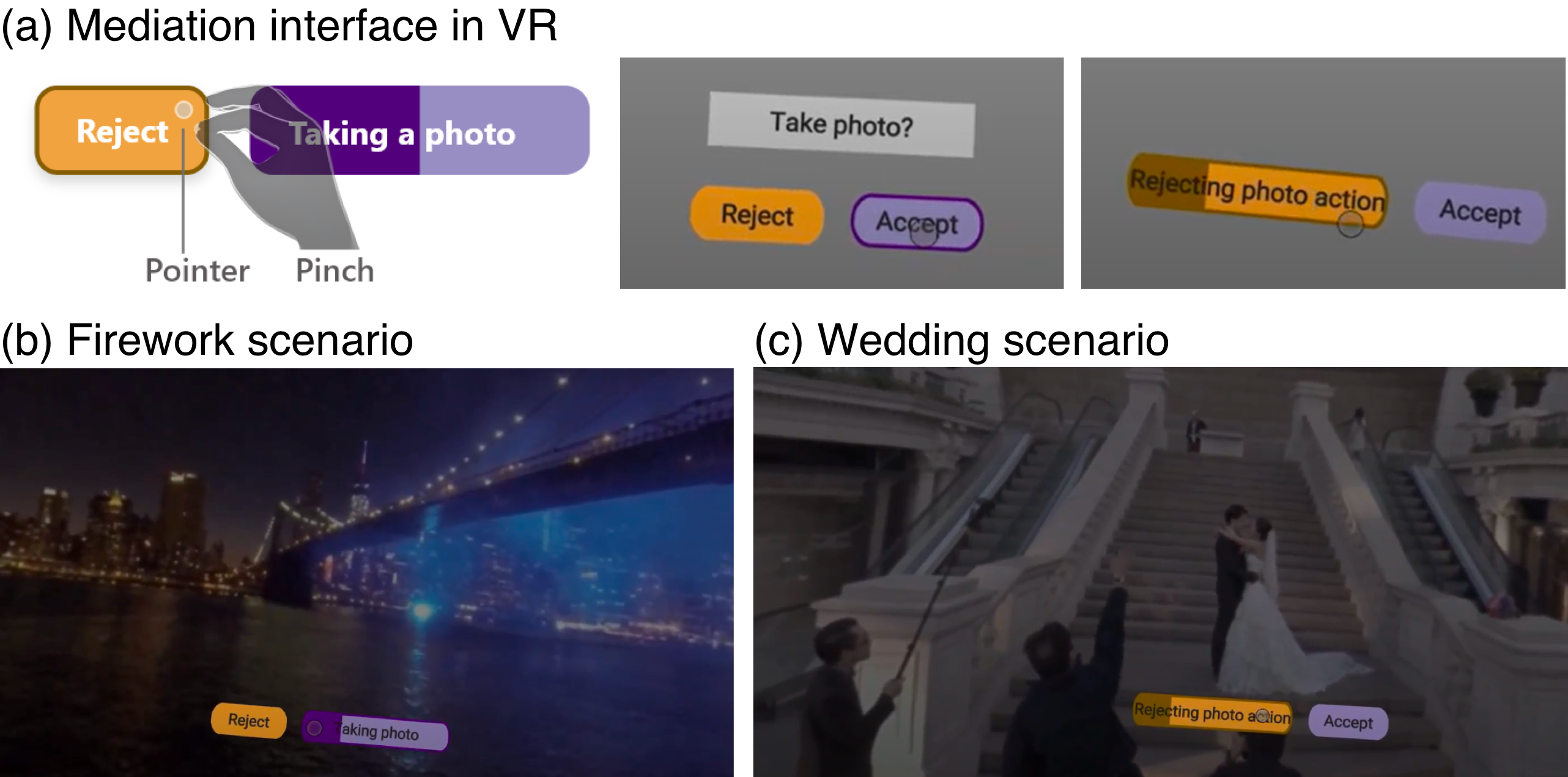}
    \vspace{-6mm}
    \caption{(a) The gesture-based mediation interface in VR. Pointing and pinching gestures were used to activate the button. (b) and (c) show the study scenarios implemented with immersive 360 videos.} 
    \label{fig:uidesign}
    \vspace{-4mm}
\end{figure}

\subsection{Study Design \& Procedure}
\label{sec:VR_study_cond}
\textbf{Immersive Scenarios.}
We selected the \pfireworks{} and \pwedding{} scenarios from the web-based study due to their distinct error costs and preferred mediation techniques. 
Both scenarios were replicated using immersive 360 videos, presenting a firework event on a boat and an outdoor wedding ceremony (Fig.~\ref{fig:uidesign}b, ~\ref{fig:uidesign}c). 
The VR glasses gave suggestions to take a photo and presented a mediation to the user to clarify their intent. When accepting the suggestion, a sound was played and the captured photo was displayed. 

\textbf{Mediation Techniques.}
Three mediation techniques (\ND{}, \DA{}, and \DR{}) were implemented with the gesture-based UI design (Fig.~\ref{fig:uidesign}a). The user accepted the mediation by performing a punch gesture with their right hand.. 

\textbf{Procedure.}
In the VR study, users provided feedback on three mediation techniques in an immersive setting. After a gesture-based UI tutorial showcasing all mediation modes, participants rated error costs and ranked mediation preferences for two scenarios. They later gave detailed feedback on their preferred technique via a PC. Scenarios and modes were counterbalanced using a Latin-square, creating six order groups. The study lasted around 1 hour.

\subsection{Participants and Experimental Setup}
Eighteen participants (M=7, F=10, Non-binary=1; Age: 18 to 65 with 83\% between 18 to 45) were recruited from the general population on dscout platform~\cite{dscout}. The participants' frequency of using VR/AR devices was never (2), rarely (10), occasionally (5), or a moderate amount (1). 
Seventeen participants reported they were right-handed and 1 reported that they were ambidextrous.

The study was designed to run on an Oculus Quest 2, with a resolution of 1832×1920 per eye. The scenarios were simulated using 360 VR images or videos with a resolution of 2160×1080. All instructions and UI designs were presented 4 meters in front of the participant. To reduce recognition complexity, hand tracking was only enabled for the right hand throughout the study using the native hand tracking functionality of the Oculus Quest 2. Participants were required to perform “pinch” gestures anywhere on the screen with their right hand to proceed through the study or perform other gestures as instructed by each design condition.
Due to COVID-19 pandemic, the study was run remotely and each participant was unsupervised.
The participants followed written instructions to install the application programmed in Unity~\cite{unity}.

\subsection{Data Collection and Analysis}

We collected subjective ratings about the perceived cost of FA and FR errors, preferred technique rankings, and participant feedback similar to the web survey. Due to the smaller sample size, 
we report inferential statistics for completeness while
focusing on the comparison to the web study results for interpretation.
The Friedman test was used for non-parametric rating data of three techniques and
Wilcoxon signed-rank test was used for pairwise comparisons.

\subsection{VR Study Results}
\label{sec:vr-phase1}
In the \pfireworks{} scenario, the frustration level of FR (Mdn=7, IQR=1) was higher than FA (Mdn=4, IQR=3.25),
with significant difference (\textit{Z} = 2.10, \textit{p} $<$ 0.05, \textit{r} = 0.58).
In the \pwedding{} scenario, the frustration level of FR (Mdn=7, IQR=1) and FA (Mdn=6, IQR=3.5) were both high
with no significant difference (\textit{Z} = 0.40, \textit{p} = 0.69, \textit{r} = 0.11).
The preferred technique for each scenario differed. In the \pfireworks{} scenario, there is a significant difference in the rankings between three techniques (\textit{$\chi^2$}(2) = 12.44, \textit{p} $<$ 0.005, \textit{W} = 0.24). \DA{} was ranked as the most preferred technique by the majority (56\%) and \DR{} was ranked by the least (0\%), with a significant difference in their rankings (\textit{p} $<$ 0.005).
In the \pwedding{} scenario, the results did not reveal a statistically significant difference in rankings between the techniques (\textit{$\chi^2$}(2) = 3.11, \textit{p}=0.21, \textit{W} = 0.06). \ND{} was ranked as the most preferred technique by the majority of participants (56\%) and \DA{} was ranked most preferred by the least (16.7\%).

\textbf{The perceived frustration in the \pfireworks{} and \pwedding{} scenarios was consistent with the web survey results, where the perceived cost of FA and FR errors were impacted by the scenarios rather than the task (i.e., photo-taking)}. The social stress introduced in the \pwedding{} scenario caused higher FA cost compared to the \pfireworks{} scenario. The preferred technique was also impacted by the perceived cost of errors. For a high cost FR scenario (e.g., \pfireworks{}), \DA{} was preferred, while for a high cost of FA scenario (e.g., \pwedding{}), \DA{} was disliked. One difference was the preferred technique in the \pwedding{} scenario. While the majority of participants preferred \DR{} in the web-based study, \ND{} was the most preferred in the VR study. Most participants expressed that they liked \ND{} because they felt the interface was \textit{easy to use with full control} (15) to take a photo at the right moment, while there was also the drawback of \textit{requiring manual input} (5). The difference may have come from the different sensory immersion in the immersive scenario, where tangible actions like taking a photo were easier to perceive than the feeling of being disrupted by the interface while chatting with friends. More discussion on the comparison of the web and immersive environment is presented in Sec.~\ref{sec:implications}.

Overall, the VR study corroborated the web study. The preferred mediation technique was impacted by the perceived cost of errors in the scenario. In addition, lower perceived FA errors resulted in a preference for the \DA{} technique while higher FA errors lead to a dislike of the \DA{} technique.

%% file: docs/6_discussion.tex

\section{Implications for XR Input Mediation}
\label{sec:implications}

We now discuss the implications of our findings on 
the design of XR mediation techniques and their generalizability and limitations.

\subsection{Perceived Error Costs}
Our research investigates how perceived error costs in various XR contexts impact the mediation techniques that users prefer to mitigate input recognition errors. 
The study results suggest that it is necessary to consider users' perceived costs when designing input error mediation techniques for XR devices, which ties back to our two research questions. When the same task is performed under different scenarios, it could induce different perceived costs of errors, which could consequently impact the choice of mediation techniques that should be used to clarify input errors. 
Based on the mediation techniques and scenarios explored in the studies, scenarios with higher costs associated to FA errors seemed to require mediation techniques that avoided FA with more user control, such as \ND{} and \DR{}. Scenarios with lower cost of FA errors seemed to require mediation techniques that were more efficient, such as \DA{}. 
These findings were also consistent across both the web-based study with the 2D interactive prototypes and the immersive study with bare-hand gestural input in VR.
Future research can develop algorithms that predict perceived costs of false accept and false reject errors using contextual information so systems can select appropriate mediation techniques for a given task and scenario.

\subsection{Using a Cost Matrix to Pinpoint Tasks and Scenarios}
Evaluating users' perceived error costs for all possible XR scenarios is unfeasible. Therefore, in our study,
we considered two common, real-life tasks with distinct FA and FR error costs. 
Based on our design exploration of task and context combinations in Sec.~\ref{sec:design_exploration}, we characterized a two-by-two error cost matrix where each task scenario could be mapped, and selected the two contrasting tasks of photo-taking and email-sending to use in our study.
By introducing social and temporal stress to vary the underlying scenarios, the perceived error costs for the tasks shifted, e.g., the FA cost increased in the \pwedding{} scenario due to social pressure.
While these tasks and scenarios do not cover the full spectrum of tasks possible within XR contexts, 
our task selection approach and study results can be generalized two-fold. 
First, the cost matrix can be used to map other real-life XR tasks and identify edge cases for a specific task. 
This mapping can help designers identify key scenarios to be considered when designing error mediation techniques for a particular XR task.
Second, varying the set of scenarios for the same task triggered a relative comparison of perceived costs from respondents. 
With our measurements focusing on the \textit{relative} costs between FA and FR errors, user perceptions can be compared across different scenarios of the same task.
Compared to an absolute perceived cost value for each task and scenario, which can vary largely by individuals and the presentation of the context, having a relative comparison across several scenarios enhances the applicability of the results. For example, if a user expressed higher concerns for FA costs in \pwedding{} than in \pfireworks{}, one could infer that \DA{} technique is likely to be less acceptable in \pwedding{}. 
As an initial attempt to characterize users' error perception across diverse real-life task scenarios, we considered using the cost matrix to pinpoint pivotal tasks and solicited user perceptions about a set of task scenarios in an effective way to cover large variations of real-life XR contexts. 
When designing mediation techniques, future XR application designers can build upon the cost-matrix approach to map users' perceived error costs under diverse XR contexts for a generalized application or under a range of selected scenarios for a specific task.

\subsection{Improving User Experiences with XR Input Mediation}

As mediation techniques are only used when a system detects uncertain user input, in most cases, users will not see a mediation interface. When input is wrongly detected (i.e., a FA or FR error occurs), the user has to manually correct the input by performing the action again or undoing the unwanted action. When the costs of FA or FR are particularly high, the impact of an input detection error on user experiences with the XR system can be significant. For example, in our imaginary camera app, we designed the system to play a sound when taking a picture to protect users' privacy. However, this would lead to user embarrassment when unintentionally taking pictures in public, as noted by a user who experienced a FA error, \textit{``You could have your hands full and it will automatically take an inappropriate picture that violates someone's privacy.''}
In such cases, mediation can be useful. For example, a camera application can temporarily store the picture until user acceptance or rejection about the camera capture is provided. This examples illustrates the importance of XR systems clearly communicating while giving the user full control, especially as XR systems increasingly act as intelligent assistants in real-life scenarios. Future work should thus focus on tailoring mediation interfaces for specific real-world applications as each application comes with its own unique set of constraints.
Further exploration of different design facets of XR mediation interfaces is also necessary, such as input techniques (e.g., gesture, voice, touch), mediation duration, and system feedback (e.g., visual, audio, haptic).

\subsection{Simulations in Web and Immersive Environments}
To control user experiences and input error consequences across participants, we simulated task scenarios in the web survey using 2D interactive prototypes and used 360 videos in the immersive VR study. While we lost some ecological validity by using simulated scenarios, the web-based study allowed us to collect a large volume of participant responses through the crowdsourced deployment. 
To mitigate the loss from this trade-off, we further recreated the scenarios in an immersive environment to increase the fidelity of the results with a smaller number of participants. 
The error perception results from the web study were internally consistent with the results from the immersive simulation, suggesting that a user's perception of error consequences is based on a shared understanding that stems from their real-world experiences. That said,  simulating real-world scenarios with 360 immersive videos can have varying effects on feeling of presence~\cite{rupp2016effects}, and impact feelings about event consequences.
For example, the wedding scenario in the web study found \DR{} was the preferred technique, but the immersive wedding scene found \ND{} was preferred. This may be due to the different level of immersion in experiencing consequences from social pressure (e.g., a camera sound while chatting with people) compared to time pressure (e.g., missing taking a photo).
While we consider immersive VR to be an effective tool to rapidly simulate diverse XR scenarios for study purposes, future studies should leverage actual XR headsets, such as Meta Quest Pro or Apple Vision Pro, to investigate user's perceived error costs in real-life scenarios when available.

\subsection{Limitations}
The most important limitation of our work is that in the web survey, participants were asked to \textit{imagine} the scenarios, instead of running the survey with actual implementations in real-world scenarios. As such, it may have been challenging for users to fully understand what the subtitles are when not being in the actual scenarios and using actual physical XR devices. While this scenario-based design approach provides important first perspectives, real-world studies should be performed in the future to validate our findings.  Furthermore, our focus was on hand-based input with button-based UIs, and therefore the results of preferred mediation designs might not generalize to other XR UIs. As more device-specific XR interactions are designed, e.g., in-car voice interfaces or gaze-activated AR glasses, future work will need to evaluate user preferences for specific mediation interface designs. Hand and voice based input may have social acceptability issues in public locations, whereas gaze based input may be more prone to input noise, which may make it less suitable for mediation, where high certainty input is needed. Future studies could look at the trade-offs between these XR input modalities for the purpose of input mediation.

%% file: docs/7_conclusion.tex

\section{Conclusion}
As XR devices provide digital information and services through natural interaction in an always-on setting, input recognition errors from hand-based input can occur more frequently and result in costly consequences. 
The results of two user studies provide evidence to support that a user's preferred mediation technique to address false accept (FA) and false reject (FR) errors differed significantly based on the perceived error costs of both task types and scenarios. The preferences were driven by the perceived cost of FA errors. The \DA{} technique was preferred when the cost of FA was low and avoided when the cost of FA was high.
Our research thus provides an empirical understanding of the impact of user-perceived error costs and the subsequent design implications of error mediation techniques in diverse XR contexts. 